\documentclass[10pt,twocolumn]{article}
\usepackage[pagenumbers]{cvpr} %

\makeatletter
\@namedef{ver@everyshi.sty}{}
\makeatother

\usepackage{graphicx}
\usepackage{amsmath}
\usepackage{amssymb}
\usepackage{booktabs}
\usepackage{siunitx}
\usepackage{import}
\usepackage[accsupp]{axessibility}

\usepackage{xcolor}
\usepackage{tabularx}

\usepackage[pagebackref,breaklinks,colorlinks]{hyperref}

\usepackage[capitalize]{cleveref}
\crefname{section}{Sec.}{Secs.}
\Crefname{section}{Section}{Sections}
\Crefname{table}{Table}{Tables}
\crefname{table}{Tab.}{Tabs.}

\def\cvprPaperID{1878} %
\def\confName{CVPR}
\def\confYear{2023}

\usepackage[firstpage=true]{background}

\SetBgContents{\parbox{\textwidth}{\footnotesize © 2023 IEEE. THIS WORK HAS BEEN SUBMITTED TO THE IEEE FOR POSSIBLE PUBLICATION. COPYRIGHT MAY BE TRANSFERRED WITHOUT NOTICE, AFTER WHICH THIS VERSION MAY NO LONGER BE ACCESSIBLE.}}
\SetBgScale{1}
\SetBgAngle{0}
\SetBgPosition{current page.south}
\SetBgVshift{1cm}
\SetBgColor{black}
\SetBgOpacity{1}

\newcommand{\LJSDE}{\mbox{L-JSDE}}
\newcommand{\RLJSDE}{\mbox{L-JSDE}}

\begin{document}

\title{Image Super-Resolution Using T-Tetromino Pixels}

\author{Simon~Grosche, 	Andy~Regensky, J\"urgen~Seiler, and~Andr\'e~Kaup\\
Chair of Multimedia Communications and Signal Processing,\\ Friedrich-Alexander-Univerist\"at Erlangen-N\"urnberg,\\
Cauerstr. 7, 91058 Erlangen, Germany\\
{\tt\small simon.grosche@fau.de, andy.regensky@fau.de, \mbox{juergen.seiler@fau.de},  \mbox{andre.kaup@fau.de}}
}
\maketitle

\begin{abstract}
	For modern high-resolution imaging sensors, pixel binning is performed in low-lighting conditions and in case high frame rates are required. To recover the original spatial resolution, single-image super-resolution techniques can be applied for upscaling.
	To achieve a higher image quality after upscaling, we propose a novel binning concept using tetromino-shaped pixels.
	It is embedded into the field of compressed sensing and the coherence is calculated to motivate the sensor layouts used.
	Next, we investigate the reconstruction quality using tetromino pixels for the first time in literature.
	Instead of using different types of tetrominoes as proposed elsewhere, we show that using a small repeating cell consisting of only four T-tetrominoes is sufficient. For reconstruction, we use a locally fully connected reconstruction (LFCR) network as well as two classical reconstruction methods from the field of compressed sensing.
	Using the LFCR network in combination with the proposed tetromino layout, we achieve superior image quality in terms of PSNR, SSIM, and visually compared to conventional single-image super-resolution using the very deep super-resolution (VDSR) network. For PSNR, a gain of up to \SI[retain-explicit-plus]{+1.92}{dB} is achieved.
\end{abstract}

\section{Introduction}
\begin{figure}[t]
	\footnotesize
	
	\def\svgwidth{\linewidth}
	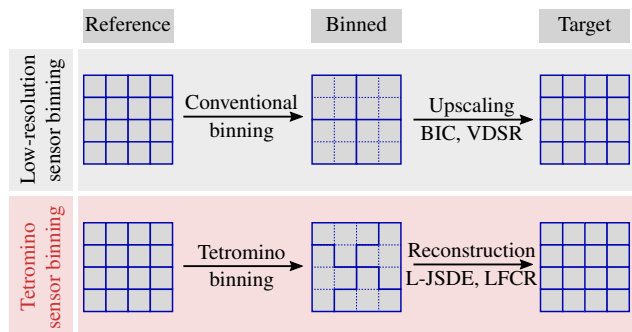
	\caption{Illustration of a conventional low-resolution binning process and the proposed tetromino binning. In both cases, less noise and higher frame rates are possible. }\label{fig:different_pixel_layouts}
\end{figure}

Conventional imaging sensors acquire image data using square pixels that are regularly placed on the sensor surface. Due to the everlasting pursuit for higher resolution, smaller and smaller pixels are packed onto the sensor surface. Such sensors can acquire single images of extremely high-resolution in case of good lighting conditions. However, there are two major issues with decreasing the pixel size. Firstly, fewer photons arrive at each pixel such that photometric limits are approaching \cite{Brueckner2013, Schoeberl2012} resulting in worse signal to noise ratios. Secondly, the required bandwidth increases with the number of measured pixels such that enormous amounts of data need to be processed and stored, especially when recording high frame rate raw videos. In order to achieve higher signal-to-noise ratios and higher frame rates, pixel binning is typically applied on the hardware level \cite{Francis2017,Janesick2001} to the disadvantage of spatial resolution. This is shown in Figure\,\ref{fig:different_pixel_layouts}.

To increase the spatial resolution of a low-resolution image, single-image super-resolution can be applied in post-processing. In this field, a vast amount of research has been investigating classical upscaling algorithms, e.g., \cite{Park2003, Freeman2002, Yang2010,  Wei2016, Wei2019}, and recently neural networks, e.g.,  \cite{Dong2016, Kim2016, Zhang2018_RDN}.  More generally speaking, the aim is to achieve the best possible image quality for a limited number of measurements.

Single-image super-resolution, however, is intrinsically limited by the regularity of the underlying measurement process which introduces aliasing whenever too high frequencies are present in the scene. One solution to circumvent artifacts from aliasing is to employ a non-regular placement of pixels \cite{Dippe1985, Hennenfent2007, Maeda2009, Anderson2013, Seiler2015, Kovarik2016}. This allows for higher image quality after reconstructing the image on a higher resolution grid without increasing the number of pixels compared to a (binned) low-resolution sensor. Unfortunately, implementations of non-regular sampling such as quarter sampling \cite{Schoberl2011} and three-quarter sampling \cite{Seiler2018} suffer from having a lower fill factor leading to stronger noise in case of low light scenarios.

To achieve a fill factor of 100\% and at the same time keep the non-regularity of the sampling process, the square shape of the conventionally used pixels must be reconsidered. For this reason, Ben-Ezra~et~al. suggest the usage of Penrose pixels to tile the sensor area \cite{Ben-Ezra2011}.
Such Penrose tiling is aperiodic, which can be understood as non-regularity on a larger scale. On the other hand, this aperiodicity also means that hardware manufacturing and readout strategies are expected to be highly complicated. %
Another possibility to achieve a 100\% fill factor with non-square pixels is the usage of hexagonal pixels \cite{Hoedlmoser2009, Vedantham2016} or triangular and rectangular shaped pixels \cite{Shi2014,Sugathan2014}.
While all these pixel shapes have potential at their own, they cannot be used for the previously described binning process within a higher resolution sensor. 
\begin{figure}[t]
	\footnotesize
	
	\def\svgwidth{\linewidth}
\begingroup%
  \makeatletter%
  \providecommand\color[2][]{%
    \errmessage{(Inkscape) Color is used for the text in Inkscape, but the package 'color.sty' is not loaded}%
    \renewcommand\color[2][]{}%
  }%
  \providecommand\transparent[1]{%
    \errmessage{(Inkscape) Transparency is used (non-zero) for the text in Inkscape, but the package 'transparent.sty' is not loaded}%
    \renewcommand\transparent[1]{}%
  }%
  \providecommand\rotatebox[2]{#2}%
  \newcommand*\fsize{\dimexpr\f@size pt\relax}%
  \newcommand*\lineheight[1]{\fontsize{\fsize}{#1\fsize}\selectfont}%
  \ifx\svgwidth\undefined%
    \setlength{\unitlength}{251bp}%
    \ifx\svgscale\undefined%
      \relax%
    \else%
      \setlength{\unitlength}{\unitlength * \real{\svgscale}}%
    \fi%
  \else%
    \setlength{\unitlength}{\svgwidth}%
  \fi%
  \global\let\svgwidth\undefined%
  \global\let\svgscale\undefined%
  \makeatother%
  \begin{picture}(1,0.40239044)%
    \lineheight{1}%
    \setlength\tabcolsep{0pt}%
    \put(0,0){\includegraphics[width=\unitlength,page=1]{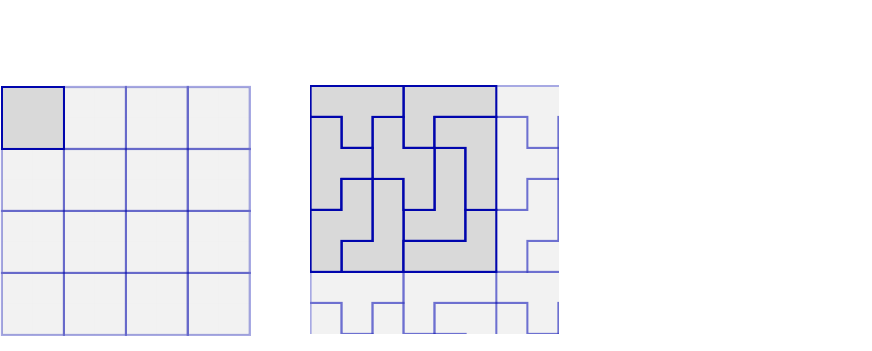}}%
    \put(-0.00172112,0.37125112){\color[rgb]{0,0,0}\makebox(0,0)[lt]{\lineheight{1.25}\smash{\begin{tabular}[t]{l}(a) Low-resolution\end{tabular}}}}%
    \put(0.04466638,0.32643038){\color[rgb]{0,0,0}\makebox(0,0)[lt]{\lineheight{1.25}\smash{\begin{tabular}[t]{l}sensor\end{tabular}}}}%
    \put(0.70307675,0.37125112){\color[rgb]{0,0,0}\makebox(0,0)[lt]{\lineheight{1.25}\smash{\begin{tabular}[t]{l}(c) $4{\times}4$ T-tetromino\end{tabular}}}}%
    \put(0.74946423,0.32643039){\color[rgb]{0,0,0}\makebox(0,0)[lt]{\lineheight{1.25}\smash{\begin{tabular}[t]{l}sensor (prop.)\end{tabular}}}}%
    \put(0,0){\includegraphics[width=\unitlength,page=2]{tetris_sensor_skizze_v2.pdf}}%
    \put(0.35477968,0.37125112){\color[rgb]{0,0,0}\makebox(0,0)[lt]{\lineheight{1.25}\smash{\begin{tabular}[t]{l}(b) Tetromino\end{tabular}}}}%
    \put(0.4011671,0.32643039){\color[rgb]{0,0,0}\makebox(0,0)[lt]{\lineheight{1.25}\smash{\begin{tabular}[t]{l}sensor from \cite{Galdo2014}\end{tabular}}}}%
    \put(0,0){\includegraphics[width=\unitlength,page=3]{tetris_sensor_skizze_v2.pdf}}%
  \end{picture}%
\endgroup%

	\caption{Illustration of different sensor layouts, pixel shapes, and cell sizes. Blue lines indicate the boundaries of the pixels. Dark gray color is used for a single complete cell. The coherence is given for an image of size $M{\times}N=30{\times}30$ pixels.}\label{fig:tetris_sensor_vs_lr_skizze}
\end{figure}

Another promising possibility are tetromino pixels as proposed in a patent application by Galdo~et\,al. \cite{Galdo2014}. In their work, T-, L- and Z-shaped tetromino pixels are used to tile the sensor area. Though it is proposed to directly manufacture the tetromino pixels in hardware, the tetromino shapes could also be used during the binning process of higher resolution pixels in case less noise or higher frame rate is desired.
Optimally, the tetromino pixels are stacked together without leaving any vacancies such that a complete tiling of the sensor area is formed. Galdo et\,al. highlight an exemplarily tiling consisting of a $6{\times}6$ pixel cell in their work which is then repeated periodically.
With respect to hardware implementation and wiring, they provide initial solutions to manufacture, connect, and read out the individual pixels. In the scope of this paper, such pixel layouts could be used for the binning of four high-resolution pixels. As for conventional square binning, this would allow for a higher signal-to-noise ratio and a higher frame rate. The resulting noise is identical for all binned sensor layouts because the light-active area is the same. Regarding the reconstruction, Galdo et\,al. suggest that techniques from compressed sensing could be used. However, no reconstruction results are given and no further analysis is provided in \cite{Galdo2014} or elsewhere. This raises the question whether a better image quality can really be achieved.

In this work, we propose a novel sensor layout based on a small  tetromino cell consisting of only four T-tetromino pixels. This layout could be used for binning four higher resolution pixels as shown in the bottom row of Figure\,\ref{fig:different_pixel_layouts}. For the proposed sensor layout as well as for other sensor layouts, we perform image reconstruction with suitable classical and data-driven algorithms. For the best of our knowledge, it is the first time in literature that image reconstruction is performed for tetromino sensor layouts. We show that the proposed sensor layout outperforms the more complicated sensor layout from \cite{Galdo2014} as well as another larger T-tetromino cell in terms of reconstruction quality.
Moreover, our tetromino sensor layout significantly outperforms the reconstruction quality of a (binned) low-resolution sensor in combination with single-image super-resolution using the very deep super-resolution (VDSR) network \cite{Kim2016}. At the same time, we are able to achieve a faster reconstruction while still outperforming VDSR.

This paper is organized as follows: In Section\,\ref{sec:lr_vs_tetromino-sampling}, we review the different sensor layouts in a compressed sensing framework. In Section\,\ref{sec:prop_T-tetrominoes}, the proposed T-tetromino sensor layout as well as a larger T-tetromino sensor layout are presented.
In Section\,\ref{sec:reconstruction_algorithms}, the used reconstruction algorithms are presented.
In Section\,\ref{sec:simulation_and_results}, we evaluate the performance of the different sensor layouts and reconstruction algorithms.
Finally, Section\,\ref{sec:conclusion} concludes the paper.

\begin{figure}[t]
	\footnotesize
	\def\svgwidth{\linewidth}
	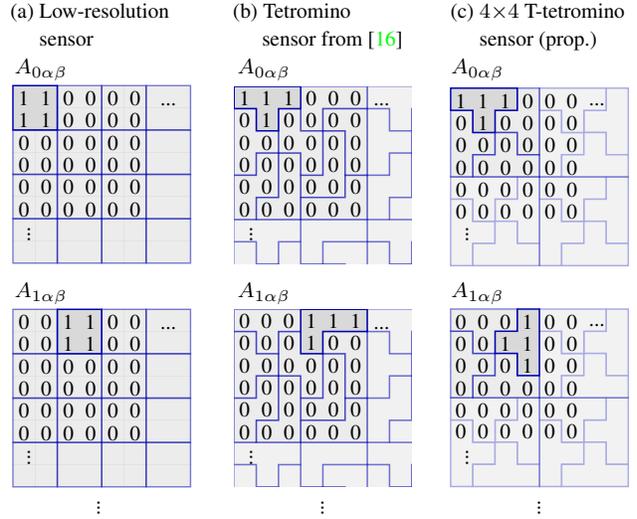
	\caption{Two slices ($i=0$ and $i=1$) through the measurement matrices for the (a) low-resolution sensor, (b) Tetromino sensor from \cite{Galdo2014}, and (c) the proposed  $4{\times}4$ T-tetromino sensor.}\label{fig:measurement_matrices}
\end{figure}

\section{Compressed Sensing Description of Conventional Sensor Layouts}
\label{sec:lr_vs_tetromino-sampling}
We describe the measurement processes for any used sensor layout within the compressed sensing framework \cite{Candes2007image, Donoho2006}.
Any compressed sensing measurement can be written as a linear combination
\begin{align}\label{eq:yi_from_A_and_f}
	y_i = \sum_{\alpha = 0}^{M-1} \sum_{\beta = 0}^{N-1} A_{i\alpha\beta} f_{\alpha\beta},
\end{align}
where $f_{\alpha\beta}\,{\in}\,[0,1]$ are the gray values of the reference image~$\boldsymbol{f}$ of size $M{\times}N$ that would be acquired with a higher resolution sensor, $y_i$ are the values measured by the individual pixels of the image sensors and $A_{i\alpha\beta}$ are the coefficients of the image measurement matrix. The index $i\,{\in}\, \{0,\dots, L-1\}$ enumerates the $L$ measurements. Moreover, $\alpha$ and $\beta$ are the vertical and horizontal positions of the pixels of the hypothetical reference image with the origin positioned in the upper left corner.
In this work, $L\,{=}\,MN/4$ for all sensor layouts.
Other than for the vast majority of the compressed sensing literature, e.g., \cite{Candes2007image, Elad2010, Gan2007, MunFowler2009}, the image is not artificially vectorized following \cite{Grosche2020_localJSDE}.

As the first sensor layout, a low-resolution sensor with large, square pixels is depicted in Figure\,\ref{fig:tetris_sensor_vs_lr_skizze}\,(a). The binning process leading to each low resolution pixel can be described as the sum of four neighboring pixels of the reference image $\boldsymbol{f}$ such that for a measurement index $i$, all entries in $A_{i\alpha\beta}$ are equal to either $1$ or $0$. Figure\,\ref{fig:measurement_matrices}\,(a) shows the slices through the three-dimensional measurement matrix for the  first two measurements with indices $i=0$ and $i=1$. If desired, the resulting measurement values $y_i$ could be reordered to form an image of lower resolution, $f^{LR}_{\hat\alpha\hat\beta} = 0.25  y_i$,
using
\begin{align}
	\hat\alpha &=  i \bmod M \equiv i - \lfloor i / M \rfloor \cdot M, \\
	\hat\beta &=  i // M \equiv\lfloor i / M \rfloor,
\end{align} 
where $\lfloor \,\cdot\, \rfloor$ is the floor operation.

The tetromino sensor layout from \cite{Galdo2014} is shown in Figure\,\ref{fig:tetris_sensor_vs_lr_skizze}\,(b). It uses T-, L- and Z-tetrominoes and has a cell size of $6{\times6}$ pixels with respect to reference image $f_{\alpha\beta}$ resulting in nine measurements, i.e., the measurement ratio $9/36=0.25$ as for the low-resolution sensor. For each measurement index $i$, most entries in $A_{i\alpha\beta}$ are again $0$ expect for four entries being equal to $1$ as shown for the first two measurements in Figure\,\ref{fig:measurement_matrices}\,(b).

Next, we motivate the potential superiority of a tetromino sensor layout by calculating the coherence $\mu$~\cite{Donoho2001, Tropp2006, Foucart2013} of the corresponding measurement matrix $A'_{i_\sigma\rho}$,
\begin{align}
	A'_{i\sigma\rho} = \sum_{\alpha = 0}^{M-1} \sum_{\beta = 0}^{N-1} A_{i\alpha\beta}  \mathit{\Phi}_{\alpha\beta \sigma\rho},
\end{align}
which is applied in the sparse transform domain $\Phi$.
The coherence $\mu$ is defined as
\begin{align}
	\mu = \max_{(\sigma,\rho)\ne(\tilde\sigma,\tilde\rho) } \left(\frac{\left|\sum_{i=0}^{L-1}A'_{i\sigma\rho}{\left(A'_{i\tilde\sigma\tilde\rho}\right)}^*\right|}{\sqrt {\sum_{i=0}^{L-1}|A'_{i\sigma\rho}|^2} \sqrt{\sum_{i=0}^{L-1}|A'_{i\tilde\sigma\tilde\rho}|^2}}\right).
\end{align}
In general, a small coherence is assumed to be beneficial~\cite{Foucart2013, Eldar2012}.
For $L\le MN$, the coherence is bound by $\mu_\mathrm{Welch} \le \mu \le 1$ with
\begin{align}
	\mu_\mathrm{Welch} = \sqrt \frac{(MN - L)}{L (MN-1)} \le \mu \le 1
\end{align}
known as the first Welch bound~\cite{Welch1974}.
\begin{table}[t]
	\caption{Coherence $\mu$ for different sensor layouts assuming an image size of $30{\times}30$ pixels.}
	\label{tab:coherence}
	\centering
	\small
	\setlength{\tabcolsep}{3pt}
	\begin{tabular}{l|c}
		 & Coherence $\mu$ \\\hline
		 Low-resolution
sensor & 1.00\\
		 Tetromino
sensor from \cite{Galdo2014} & 0.78 \\
		 4×4 T-tetromino
sensor (prop.) & 0.93
	\end{tabular}
\end{table}
The values of the coherence are given in Table\,\ref{tab:coherence} for the sensor layouts from Figure\,\ref{fig:tetris_sensor_vs_lr_skizze} assuming an image size of $30{\times}30$ pixels in order to keep the computational complexity low.
For these settings, $\mu_\mathrm{Welch} \approx 0.058$.
We find that the coherence of the low-resolution sensor is equal to one being the maximum value whereas the coherence of the tetromino sensor is smaller than one. This is an indication that using a tetromino sensor could be advantageous. Whether this is really the case needs to be found experimentally.

\section{Novel T-tetromino Sensor Layouts}
\label{sec:prop_T-tetrominoes}
In  Figure\,\ref{fig:tetris_sensor_vs_lr_skizze}\,(c) the proposed tetromino cell of size $4{\times}4$ pixels is depicted which uses only T-tetrominoes. This sensor layout is less complex than the one in Figure\,\ref{fig:tetris_sensor_vs_lr_skizze}\,(b) since the cell size is smaller and only one type of tetrominoes is used. Consequently, hardware manufacturing and wiring is expected to be simplified. Exemplarily, the first two slices through the measurement matrix are shown in Figure\,\ref{fig:measurement_matrices}\,(c).

In order to investigate the impact of the cell size for T-tetromino sensor layouts, we additionally created a T-tetromino sensor layout with a larger cell size of $8{\times}8$ pixels.
Building larger non-trivial sensor layouts with T-tetrominoes by hand is hardly possible as the T-tetrominoes may not overlap each other. At the same time, the sensor layout may be periodically geared, i.e., some of the T-tetrominoes may protrude from one cell into the next cell.
One way to create a valid T-tetromino tiling is to make use of a graph abstraction called ice graphs \cite{Korn2004}.

\begin{figure}[t]
	\small
	\def\svgwidth{\linewidth}
	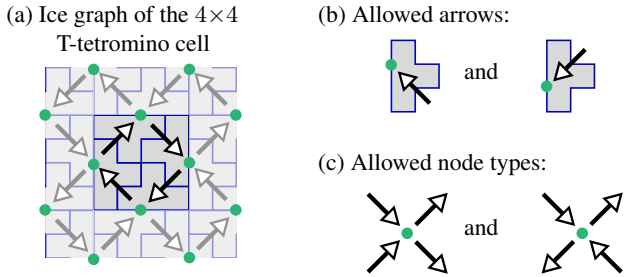
	\caption{Ice graph of the $4{\times}4$ T-tetromino sensor from Figure\,\ref{fig:tetris_sensor_vs_lr_skizze}\,(c). The rules for any valid ice graph are shown on the right side  up to rotational and mirror symmetry.}\label{fig:tetris_4x4_icegraph}
\end{figure}

Such ice graph representation of the $4{\times}4$ T-tetromino layout from Figure\,\ref{fig:tetris_sensor_vs_lr_skizze}\,(c) is depicted in Figure\,\ref{fig:tetris_4x4_icegraph}\,(a).
The ice  graph is build by connecting each node of a diagonal square lattice (green dots in Figure\,\ref{fig:tetris_4x4_icegraph}\,(a)) with its four nearest neighboring nodes. The arrow heads are then drawn centrally inside the long side of a T-tetromino pixel as shown in Figure\,\ref{fig:tetris_4x4_icegraph}\,(b). Following \cite{Korn2004}, a valid ice graph for a T-tetromino tiling has the property that the number of incoming and outgoing arrows is identical for every node. Figure\,\ref{fig:tetris_4x4_icegraph}\,(c) shows the two resulting types of nodes that are allowed up to rotation and mirror symmetry.

\begin{figure}[t]
	\small
	\def\svgwidth{\linewidth}
\begingroup%
  \makeatletter%
  \providecommand\color[2][]{%
    \errmessage{(Inkscape) Color is used for the text in Inkscape, but the package 'color.sty' is not loaded}%
    \renewcommand\color[2][]{}%
  }%
  \providecommand\transparent[1]{%
    \errmessage{(Inkscape) Transparency is used (non-zero) for the text in Inkscape, but the package 'transparent.sty' is not loaded}%
    \renewcommand\transparent[1]{}%
  }%
  \providecommand\rotatebox[2]{#2}%
  \newcommand*\fsize{\dimexpr\f@size pt\relax}%
  \newcommand*\lineheight[1]{\fontsize{\fsize}{#1\fsize}\selectfont}%
  \ifx\svgwidth\undefined%
    \setlength{\unitlength}{252bp}%
    \ifx\svgscale\undefined%
      \relax%
    \else%
      \setlength{\unitlength}{\unitlength * \real{\svgscale}}%
    \fi%
  \else%
    \setlength{\unitlength}{\svgwidth}%
  \fi%
  \global\let\svgwidth\undefined%
  \global\let\svgscale\undefined%
  \makeatother%
  \begin{picture}(1,0.54365079)%
    \lineheight{1}%
    \setlength\tabcolsep{0pt}%
    \put(0,0){\includegraphics[width=\unitlength,page=1]{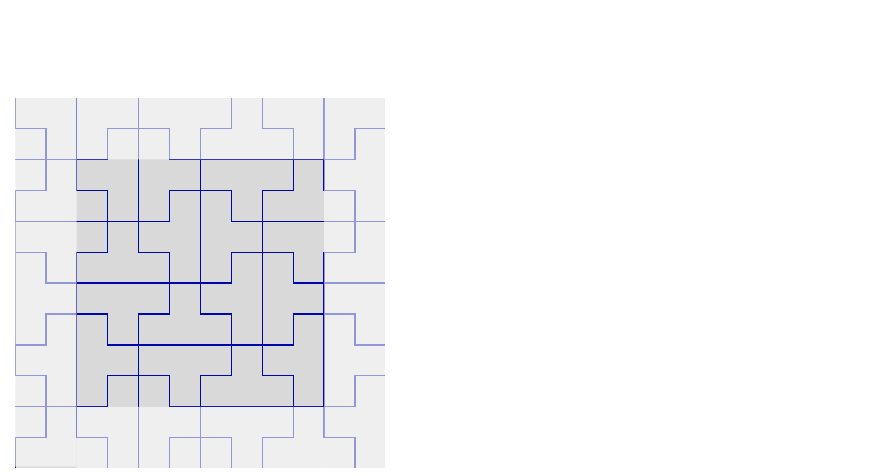}}%
    \put(0.22855397,0.50592031){\color[rgb]{0,0,0}\makebox(0,0)[t]{\lineheight{1.25}\smash{\begin{tabular}[t]{c}(a) Geared $8{\times}8$\end{tabular}}}}%
    \put(0.22832155,0.45631706){\color[rgb]{0,0,0}\makebox(0,0)[t]{\lineheight{1.25}\smash{\begin{tabular}[t]{c}T-tetromino sensor\end{tabular}}}}%
    \put(0,0){\includegraphics[width=\unitlength,page=2]{tetris_8x8_interleaved.pdf}}%
    \put(0.77542549,0.50592031){\color[rgb]{0,0,0}\makebox(0,0)[t]{\lineheight{1.25}\smash{\begin{tabular}[t]{c}(b) Corresponding\end{tabular}}}}%
    \put(0.77572311,0.45631706){\color[rgb]{0,0,0}\makebox(0,0)[t]{\lineheight{1.25}\smash{\begin{tabular}[t]{c}ice graph\end{tabular}}}}%
    \put(0,0){\includegraphics[width=\unitlength,page=3]{tetris_8x8_interleaved.pdf}}%
  \end{picture}%
\endgroup%

	\caption{Ice graph of a randomly generated $8{\times}8$ T-tetromino sensor. Other than the previous tetromino tilings, this tiling itself does not show any cell boundaries since the pixels are tightly geared into each other after periodic repetition.}\label{fig:tetris_8x8_icegraph_interleaved}
\end{figure}

Using this ice graph representation, $2^{((8/2)^2)}= 2^{16}$ different graphs are in principle possible for a cell size of $8{\times}8$ pixels. We therefore randomly generated roughly $2^{16}$ graphs for a cell size of $8{\times}8$ pixels and tested which graphs are valid with respect to the allowed node types in Figure\,\ref{fig:tetris_4x4_icegraph}\,(c). The first valid ice graph among those randomly generated graphs is depicted in Figure\,\ref{fig:tetris_8x8_icegraph_interleaved} together with its corresponding T-tetromino tiling. %
For larger cell sizes such as 16 or 32, this brute-force approach was not successful within reasonable computing time, because the number of possible graphs increases exponentially. Generating larger tilings requires more sophisticated approaches, see e.g. \cite{Bodini2010}, but is neglected in this work as the results for the smaller cell size will be even superior than those for the larger cell sizes.

\label{sec:reconstruction_algorithms}

After the measurements have been performed, the image needs to be reconstructed as illustrated in Figure\,\ref{fig:different_pixel_layouts}. The reconstruction algorithm has to find a solution $\hat{f}_{\alpha\beta}$ from the measurements $y_i$ such that the measurement equation (\ref{eq:yi_from_A_and_f}) is satisfied as good as possible, e.g., in a least square sense.
For measurements from the low-resolution sensor (Figure\,\ref{fig:tetris_sensor_vs_lr_skizze}\,(a)), we use bicubic upscaling (BIC) \cite{Walt2014} as well as the very deep super-resolution (VDSR) network \cite{Kim2016} which is a widely used neural network for super-resolution. In order to match the sensor layout in Figure\,\ref{fig:tetris_sensor_vs_lr_skizze}\,(a), we use a binning of four high-resolution pixels whereas the original publication of VDSR uses bicubic downscaling. Consequently, we retrained the model for the low-resolution sensor. Otherwise, the results of VDSR would be significantly worse due to the mismatch of the downscaling procedures \cite{Koehler2020}.

For the tetromino sensor layouts in Figure\,\ref{fig:tetris_sensor_vs_lr_skizze}\,(b,c) and Figure\,\ref{fig:tetris_8x8_icegraph_interleaved}, the reconstruction algorithms should make use of the spatial extend of each pixel.
We apply two classical reconstruction algorithms from the field of compressed sensing that are capable of reconstructing images from a broad class of measurement matrices. They make use of the spatial extend of each pixel by explicitly exploiting the  measurement matrix, $A_{i\alpha\beta}$. Moreover, we employ a neural network proposed for other non-regular sampling tasks. This data-driven approach implicitly learns the shape and position of the pixels during the training process.
\begin{figure*}[t]
	\small
	\def\svgwidth{\textwidth}
	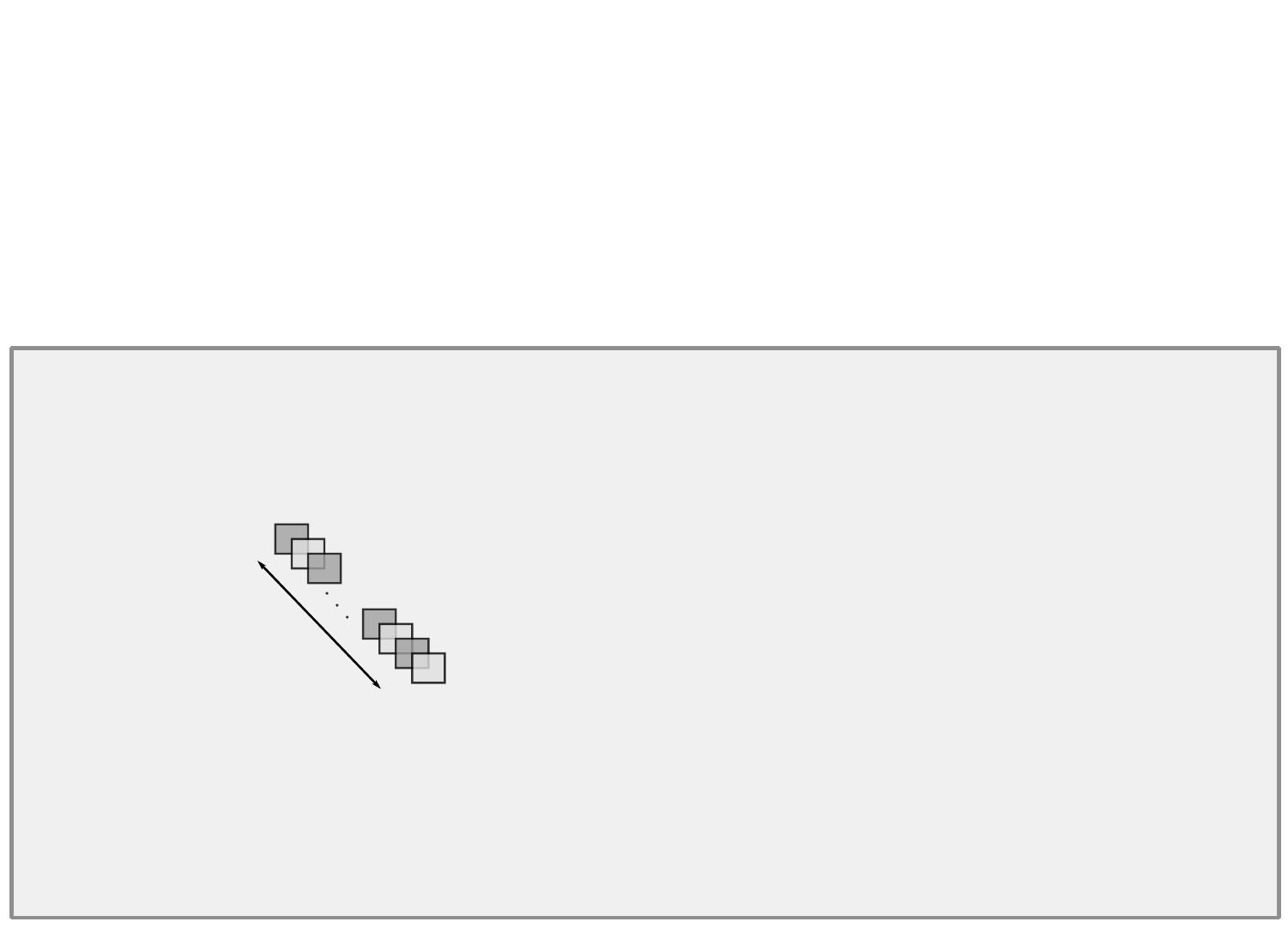
	\caption{Illustration of the LFCR network \cite{Grosche2021_LFCR} for case of the proposed $4{\times}4$ T-tetromino sampling sensor.}\label{fig:LFCR_for_4x4_Tetris}
	\vspace*{-0.5cm}
\end{figure*}
\section{Reconstruction Algorithms}

As the first  general compressed sensing algorithm, we use local joint sparse deconvolution and extrapolation (\mbox{L-JSDE}) \cite{Grosche2020_localJSDE}. It was proposed for the reconstruction of images from arbitrary local measurements.  \mbox{L-JSDE} is an extension of JSDE \cite{Seiler2018} which was first designed to only reconstruct  images taken with a three-quarter sampling sensor \cite{Seiler2018}. \mbox{L-JSDE} is based on an overlapping sliding window approach and generates an iterative model in the discrete Fourier transform domain for each model window. We use a target block size of $B{\times}B\,{=}\,4{\times}4$ pixels and a model window of size $W{\times}W\,{=}\,32{\times}32$ pixels. This is consistent with the choices for similar tasks in \cite{Grosche2020_localJSDE}. All other parameters are chosen as in  \cite{Grosche2020_localJSDE}, too. An implementation is provided by the authors.

The second general reconstruction algorithm is based on a smoothed projected Landweber (SPL) iteration and was first suggested in \cite{Gan2007} for block-wise compressed sensing problems.
We use an implementation provided by the authors \cite{MunFowler2009} with their default settings. It uses a dual-tree discrete wavelet transform \cite{Kingsbury2001} as basis functions.
For each block, an iterative projected Landweber reconstruction is performed. In each iteration, a projection step is followed by a thresholding step.
Then, the blocks are re-combined for a joint Wiener filtering \cite{lim1990two} with a $3{\times}3$ kernel before the next Landweber iteration is performed within the blocks.
The Wiener filtering prevents blocking artifacts at the borders of the blocks and at the same time reduces noise.
For the tetromino sensor layouts, other than for block-wise compressed sensing, we cannot decompose the reconstruction problem into neighboring reconstruction blocks. This is the case because the cells of the tetromino sensor layouts may be geared at the boundary of the reconstruction block (cf. Figure\,\ref{fig:tetris_8x8_icegraph_interleaved}).
Decomposing such measurements into neighboring blocks would therefore neglect some of the measurements at the block boundaries which effectively reduces the measurement rate. To solve this issue occurring with SPL, we extend SPL from \cite{MunFowler2009} with an overlapping sliding window approach similar to the one present in \cite{Grosche2020_localJSDE}.
We use neighboring target blocks of size  $16{\times}16$ pixels and  a model window size of $32{\times}32$ pixels. Each model window is then reconstructed separately. Since only the central target blocks of the reconstructed model windows are used for the final image, blocking artifacts are effectively prevented and all measurements can contribute to the overall image reconstruction.

For the data-driven reconstruction algorithm, we rely on a recent neural network based approach, namely the locally fully connected  reconstruction (LFCR) network \cite{Grosche2021_LFCR}. It finds a reconstruction for a block of size $8{\times}8$ pixels from all measurements positioned inside this block as well as those within a border of 4 pixels. The LFCR network is proposed for cases such as quarter sampling and three-quarter sampling but can also be applied for the low-resolution sensor \cite{Grosche2021_LFCR}. 
Since the LFCR assumes  a periodicity of $8{\times}8$ pixels for the measurement process, the tetromino pattern from \cite{Galdo2014} (cf. Figure\,\ref{fig:tetris_sensor_vs_lr_skizze}\,(b)) is not used with LFCR as this would require to change the entire network structure and the number of trainable parameters. This would make a fair comparison hardly possible.
The LFCR itself is a convolutional neural network. For the first convolution, a stride of 8 is used which enforces a  translational invariance after a displacement of 8 pixels. This convolution is non-trainable and simulates the sensor behavior. The entries of this convolution are set such that the T-tetromino sensor layouts are resembled. Next, several convolutions with a kernel size of $1{\times}1$ establish a fully connected link between the measured data and the reconstructed pixels values. Finally, a de-convolution (also called transposed convolution) with stride 8 is performed to re-order the reconstructed pixels to the a full image.
The LFCR network is concatenated with a VDSR-like network consisting of twenty  convolutional layers with a kernel size of $3{\times}3$ and a residual connection. Other than in the original publication of VDSR \cite{Kim2016}, we use the  PReLU activation function \cite{He2015_PReLu} for LFCR as well as the concatenated VDSR. As in \cite{Grosche2021_LFCR}, the LFCR+VDSR network is trained in a two-step procedure. First, only the LFCR is trained on the mean squared error with respect to the high-resolution reference. Next, only the second half (i.e., the  VDSR) is trained using the same loss function.

\section{Experiments and Results}
\label{sec:simulation_and_results}
\subsection{Simulation Setup}
We performed experiments to evaluate the reconstruction quality using the tetromino sensor layouts compared to the low-resolution sensor using different reconstruction algorithms.
For the reconstruction algorithms, we use the parameters  as detailed in Section\,\ref{sec:reconstruction_algorithms} following the respective literature. 
In order to train the neural networks, LFCR, LFCR+VDSR and VDSR, we use the Set291 dataset as in the original publication of VDSR \cite{Kim2016}.
Regarding the evaluation dataset, we use the TECNICK image dataset \cite{Asuni2014} consisting of 100 natural images of size $1200{\times}1200$ pixels and the Urban100 dataset \cite{Huang2015} consisting of 100 images in the megapixel range, too. 
The images serve as reference images $\boldsymbol{f}$ and the measured values $y_i$ can be generated by multiplying the image with the measurement matrix of the respective sensor layouts  as in (\ref{eq:yi_from_A_and_f}).

\begin{figure}[t]
	\footnotesize
	\def\svgwidth{\linewidth}
	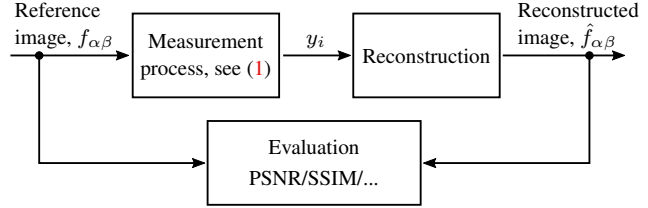
	\caption{Overview of the processing pipeline for the simulations setup. For the evaluation, the reference image is compared to the reconstructed image.}\label{fig:blockschaltbild_PSNR}
\end{figure}

Subsequent to the reconstruction, the peak signal-to-noise ratio (PSNR) is calculated with respect to the respective reference image
\begin{align}
	\mathit{PSNR} = 10\log_{10} \frac{255^2}{\frac{1}{M N}\sum_{\alpha\beta} (f_{\alpha_\beta} - \hat{f}_{\alpha\beta})^2}
\end{align}
and is then averaged for all 100 images of the datasets. As common in the literature, a border of 16 pixels is neglected for the PSNR calculations to avoid potential boundary effects.
As another metric, the widely used structural similarity index measure (SSIM)~\cite{Wang2004} is evaluated in the same way.
The full processing chain for the simulations is summarized in Figure\,\ref{fig:blockschaltbild_PSNR}.

\begin{figure*}[t]
	\def\svgwidth{\textwidth}
	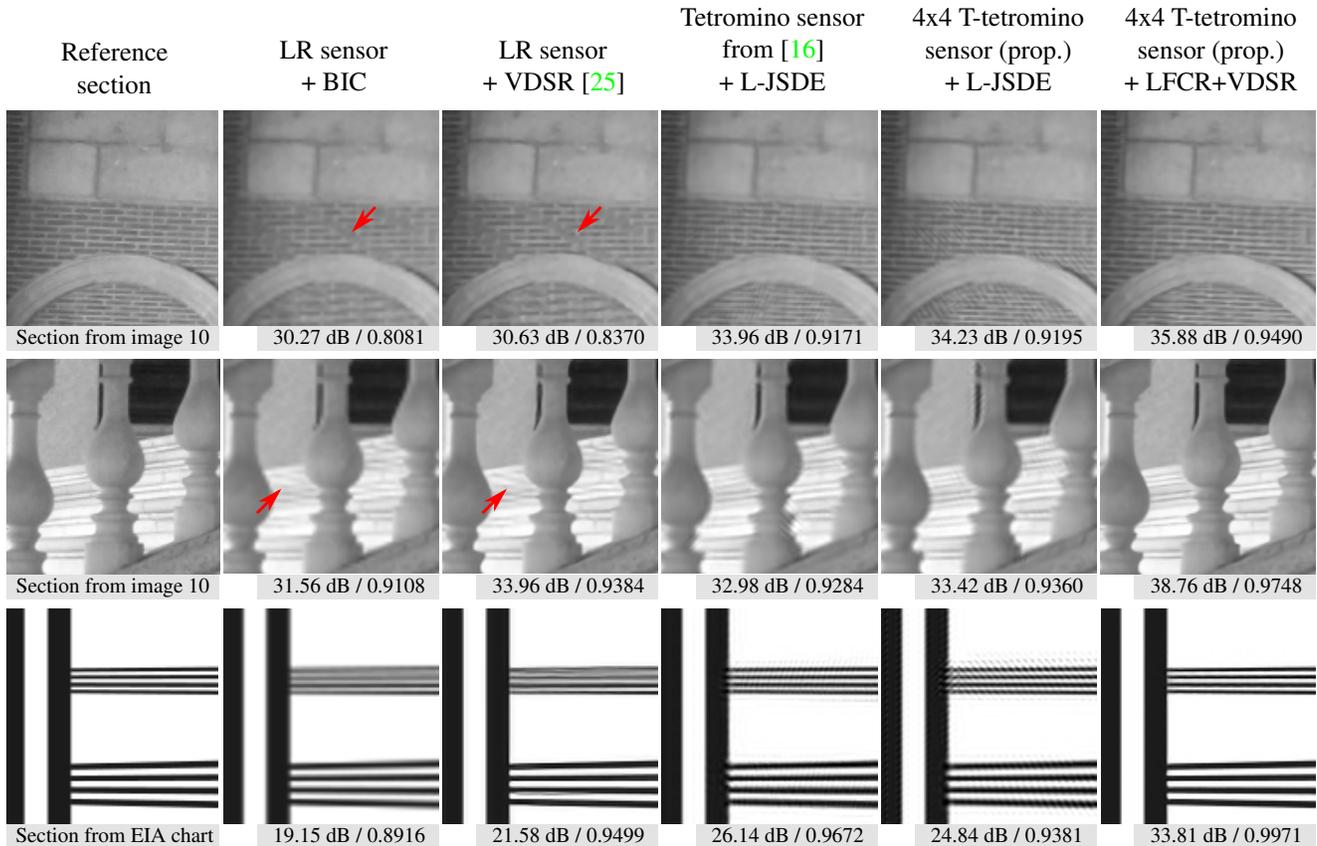
	\caption{Visual comparison of sections where single-image super-resolution fails as a consequence of the aliasing introduced by the low-resolution (LR) sensor. \textit{(Please pay attention, additional aliasing may be caused by printing or scaling. Best to be viewed enlarged on a monitor. The reference images do not show visible aliasing artifacts.)}}\label{fig:results_vs_vdsr}
	\label{fig:compare_VDSR}
	\vspace*{-0.5cm}
\end{figure*}

\subsection{Comparison of the Different Tetromino Sensor Layouts}
\begin{table}[t]
	
	\caption{Reconstruction quality in terms of PSNR in dB averaged for the 100 images from the TECNICK dataset. The results are shown for the three different tetromino sensor layouts and three reconstruction algorithms. The best results in each column is highlighted with bold font.}
	\label{tab:psnr_tetris_compare}
	\centering
	\small
	\setlength{\tabcolsep}{3pt}
	\begin{tabular}{l||c|c|c}
		&           \LJSDE{}           &            SPL            &         LFCR+VDSR         \\
		& \cite{Grosche2020_localJSDE} &   \cite{MunFowler2009}    &  \cite{Grosche2021_LFCR}  \\ \hline
		Tetromino sensor from \cite{Galdo2014} &       34.27       &     31.55      &             -             \\
		Geared 8x8 T-tetromino             &       34.30       &     31.71      &     36.63      \\
		4x4 T-tetromino  (prop.)           &  \textbf{34.49}   & \textbf{31.76} & \textbf{37.32}
	\end{tabular}
\end{table}
\begin{table}[t]
	
	\caption{Reconstruction quality in terms of SSIM averaged for the 100 images from the TECNICK dataset. The results are shown for the three different tetromino sensor layouts and three reconstruction algorithms. The best results in each column is highlighted with bold font.}
	\label{tab:psnr_tetris_compare_SSIM}
	\centering
	\small
	\setlength{\tabcolsep}{3pt}
	\begin{tabular}{l||c|c|c}
		&           \LJSDE{}           &            SPL            &         LFCR+VDSR         \\
		& \cite{Grosche2020_localJSDE} &   \cite{MunFowler2009}    &  \cite{Grosche2021_LFCR}  \\ \hline
		Tetromino sensor from \cite{Galdo2014} &       0.9649       &     0.9408      &             -             \\
		Geared 8x8 T-tetromino             &       0.9654       &     0.9427      &     0.9759      \\
		4x4 T-tetromino  (prop.)           &  \textbf{0.9695}   & \textbf{0.9467} & \textbf{0.9785}
	\end{tabular}
\end{table}
Table\,\ref{tab:psnr_tetris_compare} and Table\,\ref{tab:psnr_tetris_compare_SSIM} show the reconstruction quality  in terms of PSNR and SSIM using the three different tetromino sensor layouts with various reconstruction algorithms. In both tables, the average values for the TECNICK dataset are given.
The general purpose reconstruction algorithms, \LJSDE{} \cite{Grosche2020_localJSDE} and SPL \cite{MunFowler2009}, can be used for all sensor layouts, whereas LFCR+VDSR \cite{Grosche2021_LFCR} cannot be applied to the tetromino sensor layout form \cite{Galdo2014} as discussed in Section\,\ref{sec:reconstruction_algorithms}.

Among the three tetromino sensor layouts, the $4{\times}4$ \mbox{T-tetromino} sensor performs best in terms of PSNR and SSIM for all three reconstruction algorithms. Using this sensor layout results in a PSNR gain of at least 0.21\,dB compared to the tetromino sensor layout from \cite{Galdo2014}. Likewise, the SSIM is increased.
In the next section, we compare the performance of this proposed $4{\times}4$ T-tetromino sensor  with conventional single-image super-resolution.

\subsection{Comparison with Single-Image Super-Resolution}

In Table\,\ref{tab:results_VDSR_PSNR}, the average image quality in terms of PSNR and SSIM is given using the low-resolution sensor in combination with bicubic upscaling and VDSR \cite{Kim2016} in comparison to using the proposed  $4{\times}4$ T-tetromino sensor with L-JSDE as well as LFCR (only), i.e., without the concatenated VDSR network, and LFCR+VDSR \cite{Grosche2021_LFCR}. 
In addition to the TECNICK dataset (cf. Table\,\ref{tab:psnr_tetris_compare}), we show the average results for the Urban100 dataset which was also used in \cite{Kim2016}.
\begin{table}[t]
	\caption{Average PSNR in \si{dB} and SSIM values for state-of-the-art single-image super-resolution compared to reconstruction results for the proposed $4{\times}4$ T-tetrominoes sensor. average results are shown for both dataset. The best results in each column are highlighted with bold font.}
	\label{tab:results_VDSR_PSNR}
	\centering
	\setlength{\tabcolsep}{3pt}
	\small
	\begin{tabular}{l||c|c}
		& TECNICK \cite{Asuni2014}  & Urban100  \cite{Huang2015} \\ \hline
		Low-resolution sensor                                &                           &                            \\
		\quad\quad + BIC   \cite{Walt2014}                   &     33.66\,/\,0.9631      &      25.67\,/\,0.8820      \\
		\quad\quad + VDSR  \cite{Kim2016}                    &     36.20\,/\,0.9746      &      28.92\,/\,0.9299      \\
		\quad\quad + LFCR+VDSR \cite{Grosche2021_LFCR}       &     36.01\,/\,0.9739      &      28.73\,/\,0.9283      \\ \hline
		$4{\times}4$ T-tetromino   (prop.)                   &                           &                            \\
		\quad\quad + \LJSDE{}   \cite{Grosche2020_localJSDE} &     34.49\,/\,0.9695      &      26.94\,/\,0.9050      \\
		\quad\quad + LFCR (only)   \cite{Grosche2021_LFCR}   &     36.84\,/\,0.9771      &      29.59\,/\,0.9386      \\
		\quad\quad + LFCR+VDSR \cite{Grosche2021_LFCR}       & \textbf{37.32\,/\,0.9785} & \textbf{30.54\,/\,0.9475}
	\end{tabular}
\end{table}

From Table\,\ref{tab:results_VDSR_PSNR}, it can be seen that the $4{\times}4$ T-tetromino sensor with LFCR+VDSR \cite{Grosche2021_LFCR} performs better than the low-resolution sensor with VDSR \cite{Kim2016} for both image datasets. For Urban100, the gain in terms of PSNR is largest with \SI{1.62}{dB}.
The second best results are achieved when the $4{\times}4$ T-tetromino sensor is reconstructed using LFCR (only) for which the gain compared to single-image super-resolution with VDSR is \SI{0.67}{dB}.

The differences in PSNR and SSIM can be attributed to the visual results in Figure\,\ref{fig:compare_VDSR} where two sections from an image of the TECNICK dataset are shown. Additionally, a section from the EIA resolution chart 1956 is depicted.
In the first row, the fine, periodic structure of the bricks introduces aliasing when the low-resolution sensor is used. Consequently,  diagonal low-frequency stripes are visible for the bicubically upscale image. Since the artifacts are quite strong, VDSR \cite{Kim2016} suffers from the same artifacts. 
Similarly, for the second row, VDSR \cite{Kim2016} creates curvy stairs even though these should be straight. Again this can be attributed to the actual aliasing introduced in the measurement process.
This becomes even clearer in the third row where four fine lines are shown on a resolution test chart.
For all three examples, the reconstruction for the $4{\times}4$ T-tetromino sensor layout is superior and no such aliasing is visible. This is the case for both shown reconstruction algorithms though the results with LFCR+VDSR \cite{Grosche2021_LFCR} outperform those with L-JSDE \cite{Grosche2020_localJSDE} as in Tab.\,\ref{tab:results_VDSR_PSNR}. For completeness, the reconstruction result for the tetromino sensor from \cite{Galdo2014} using \LJSDE{} are also depicted. 

Overall, we find that the occurrence of visual artifacts and defects attributed to aliasing can be significantly reduced by using the tetromino sensor layouts. This advantage originates from the non-regularity of the measurement process allowing for a better image reconstruction.

\subsection{Evaluation of the Runtime}

\begin{figure}[t]
	\def\svgwidth{\linewidth}
	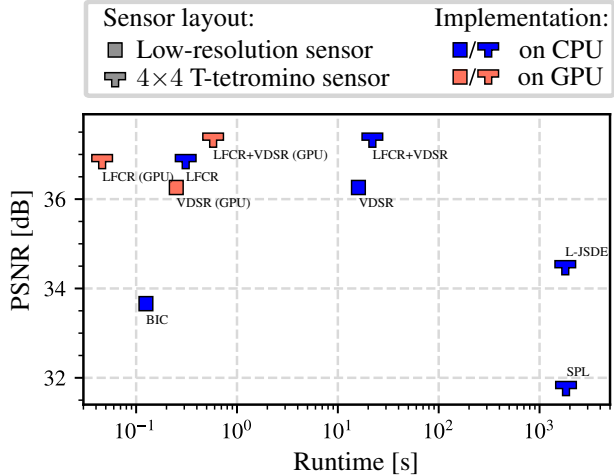
	\caption{Runtime and image quality in terms of PSNR in dB for different reconstruction algorithms and sensor layouts. The fastest reconstruction is achieved using LFCR (only) on GPU.}\label{fig:results_runtime_vs_PSNR}
	\vspace*{-0.5cm}
\end{figure}
In this section, we investigate the runtime for the different reconstruction algorithms.
All programs were restricted to a single CPU core of an \textit{Intel Xeon E3-1245v5} processor with 3.50\,GHz. The neural networks were additionally run on a \textit{Nvidia GeForce RTX 2060 SUPER} GPU. In each case, the measured runtime is averaged for the first ten images of the TECNICK dataset being of size $1200{\times}1200$ pixels.

The timing results are provided in Figure\,\ref{fig:results_runtime_vs_PSNR} for the low-resolution sensor layout as well as  $4{\times}4$ T-tetromino sensor. The runtime is presented together with the respective reconstruction qualities in terms of PSNR on the vertical axis.

As a first observation, we find that the classical algorithms are much slower than the neural network, even when the networks are executed on CPU.
Among the neural networks, the GPU version of LFCR (only), i.e., without concatenated VDSR, \cite{Grosche2021_LFCR} is the fastest being more than 5 times faster than VDSR \cite{Kim2016} on GPU. It allows for up to 20 frames per second. The complete LFCR+VDSR \cite{Grosche2021_LFCR} is slightly slower than VDSR \cite{Kim2016} as the combined network is larger. However, this network achieves the highest reconstruction quality (in combination with our $4{\times}4$ T-tetromino) and the additional processing time may be acceptable in case the highest image quality is desired.

\section{Conclusion and Future Work}
\label{sec:conclusion}
In this paper, we investigated the usage of tetromino pixels for a novel sensor layout concept for imaging sensors. Such pixels could be used in hardware binning instead of conventional $2{\times}2$ binning in order to increase the signal to noise ratio and the frame rate. For the first time in literature, we provide reconstruction results for such tetromino sensor layouts. We propose using a simple $4{\times}4$ cell consisting of only  T-tetromino which performs better than a more complicated design from literature.  Moreover, we compare our results to the case of single-image super-resolution with a (binned) low-resolution sensor in terms of image quality and runtime.

Using our $4{\times}4$ T-tetromino sensor layout in combination with general purpose algorithms from compressed sensing as well as data-driven approaches, we can show that the reconstruction quality is superior than for a larger tiling with different tetromino pixels from literature as well as a larger tilings with T-tetrominoes.

Comparing the proposed $4{\times}4$ T-tetromino sensor layout with state-of-the-art neural networks for single-image super-resolution, we find that regions with strong aliasing artifacts can be reconstructed better. This is attributed to the non-regularity of the pixel shape and arrangement  being a physical advantage over using the same number of square, regularly placed pixels. This finding is also reflected in the average PSNR and SSIM results. For the PSNR, LFCR+VDSR outperforms VDSR by \SI[retain-explicit-plus]{+1.62}{dB} for the Urban100 dataset.

Regarding the runtime, LFCR (only) shows the fastest results being 5 times faster than VDSR while still outperforming the VDSR in terms of PSNR. In case the highest image quality is desired with LFCR+VDSR, the runtime is slightly slower than for VDSR due to the larger network architecture.

For future work, it might be possible to include insights from more recent single-image super-resolution networks. These could be appended to LFCR instead of VDSR which is expected to moderately increase the reconstruction quality. Moreover, their concepts could potentially be included within LFCR, e.g., by using residual dense connections as in \cite{Zhang2018_RDN}.
Besides, more than four pixels can also be combined. Hexominoes (6 pixels) or octominoes (8 pixels) may be reasonable choices to further increase the pixel size.

{\small
\bibliographystyle{ieee_fullname}
\bibliography{literatur_jabref_backup_20220822,egbib}
}

\end{document}